\documentclass[fleqn,10pt]{SelfArx} 

\definecolor{color1}{RGB}{0,0,90} 
\definecolor{color2}{RGB}{0,20,20} 

\usepackage{hyperref} 
\hypersetup{hidelinks,colorlinks,breaklinks=true,urlcolor=color2,citecolor=color1,linkcolor=color1,bookmarksopen=false,pdftitle={Title},pdfauthor={Author},urlcolor=blue}

\usepackage{graphicx}
\usepackage[square]{natbib}
\usepackage{amsmath}
\usepackage{multirow}
\usepackage{subfigure}
\usepackage{verbatim}
\usepackage{siunitx}

\newcommand{\n}[1]{\mathrm{#1}}

\JournalInfo{Published in Journal of Electronic Materials, Vol. 45 (3), 1301-1308, 2016} 
\Archive{\href{http://dx.doi.org/10.1007/s11664-015-4014-z}{DOI: 10.1007/s11664-015-4014-z}} 

\PaperTitle{An analytical model for the influence of contact resistance on thermoelectric efficiency} 

\Authors{R. Bj\o{}rk} 
\affiliation{\textit{Department of Energy Conversion and Storage, Technical University of Denmark - DTU, Frederiksborgvej 399, DK-4000 Roskilde, Denmark}} 
\affiliation{*\textbf{Corresponding author}: rabj@dtu.dk} 

\newcommand{\keywordname}{Thermoelectrics, Efficiency, Contact resistance, Electrical resistance, Thermal resistance} 

\Abstract{An analytical model is presented that can account for both electrical and hot and cold thermal contact resistances when calculating the efficiency of a thermoelectric generator. The model is compared to a numerical model of a thermoelectric leg, for 16 different thermoelectric materials, as well as the analytical models of Ebling et. al. (2010) and Min \& Rowe (1992). The model presented here is shown to accurately calculate the efficiency for all systems and all contact resistances considered, with an average difference in efficiency between the numerical model and the analytical model of $-0.07\pm0.35$ pp. This makes the model more accurate than previously published models. The maximum absolute difference in efficiency between the analytical model and the numerical model is 1.14 pp for all materials and all contact resistances considered.}

\begin{document}

\flushbottom 

\maketitle 


\thispagestyle{empty} 

\section{Introduction}
A thermoelectric (TE) generator is limited in efficiency by intrinsic factors, such as material properties, but also by extrinsic factors such as heat loss or contact resistance. It is of equal importance to consider both the intrinsic and extrinsic factors when new scientific or commercial TE devices are designed and constructed. The intrinsic efficiency of a thermoelectric generator is determined by the material properties, through the thermoelectric figure of merit, $ZT$, which is defined as
\begin{math}
ZT = \frac{\alpha^2T}{\rho\kappa}
\end{math}
where $\alpha$ is the Seebeck coefficient, $\rho$ is the resistivity, $\kappa$ is the thermal conductivity and $T$ is the absolute temperature.

Regarding the extrinsic factors, the two most important are heat loss and contact resistance. The former has been investigated in some detail in the literature, and can be avoided to a certain degree by proper insulation \cite{Ziolkowski_2010,Bjoerk_2014}. The latter factor, contact resistance, consists of two components, namely electrical and thermal contact resistance. Experimentally, especially the manufacture of an electrical contact for thermoelectrics with a low electrical contact resistance has been studied, in order to improve the device efficiency \cite{ElGenk_2003,ElGenk_2004,ElGenk_2006,DAngelo_2011,Hung_2014}. This is important as e.g. adding electrodes to a single Mg$_2$Si leg increased the total leg resistance by a factor of 2-3 \cite{Sakamoto_2012}. So far metal contacts \cite{DAngelo_2007}, titanium disilicide (TiSi$_2$) \cite{Assion_2013}, transition-metal silicides \cite{Sakamoto_2012}, silver-based alloys \cite{Gan_2013}, antimony (Sb) \cite{Li_2012}, Ti foil \cite{Zhao_2012} and Ag and Cu \cite{Zybala_2010} contacts have been tried. Typical values of the specific electrical contact resistance are $\sim{}10^{-5}$ $\mathrm{\Omega}$ cm$^2$ \cite{DAngelo_2007,Li_2012}. However, these experimental studies have not considered the actual influence of the contact resistance on the performance of the TE device, but only sought to minimize the resistance. For the case of a thermal contact resistance, the resistance will lower the temperature span across the device. This will cause a decrease in efficiency, but this decrease will depend on the material properties and is not known analytically.

Numerically, only the influence of electrical contact resistance has been considered, and this only for non-segmented bismuth telluride leg \cite{Pettes_2007, Ebling_2010,Reddy_2014}. For segmented legs, the influence of electrical and thermal contact resistance on the efficiency has been investigated numerically \cite{Bjoerk_2015}, for a wide variety of TE materials. Here, a universal influence of both the electrical and thermal contact resistance on the leg's efficiency was observed when the systems were analyzed in terms of the contribution of the contact resistance to the total resistance of the device.

Here, we consider the influence of a thermal or electrical contact resistance on the efficiency of a single thermoelectric leg, consisting of one material. At present, two analytical models capable of describing the influence of contact resistance on the performance of the thermoelectric device exists. One is the model by Min \& Rowe \cite{Min_1992,Rowe_1996,TE_Handbook_Ch9}, which considers both an electrical and thermal contact resistance. While this model seems to completely describe the relevant physics for thermoelectric devices, previous work has shown the model to be inaccurate for segmented legs \cite{Bjoerk_2015}. This has allowed for speculation as to whether the predictions of the Min \& Rowe model is accurate for non-segmented legs as well. This will be investigated in this work. The other existing analytical model, by Ebling et. al. \cite{Ebling_2010}, considers only an electrical contact resistance, and has so far not been compared to numerical predictions for a range of thermoelectric materials. Furthermore, this model assumes a change of the $ZT$ value of a material when an electrical contact resistance is present. This seems unphysical, as an external parameter, such as contact resistance, cannot change an internal parameter, such as the material parameter $ZT$. Finally, both of these existing analytical models are presented without a thorough derivation in their representative publications. This makes it hard to understand the underlying physical assumptions for both models.

Here, we will consider a new analytical model that can account for the influence of contact resistance and accurately calculate the efficiency of a TE leg. The model will be based on clear physical arguments and will be derived in as thorough and clear manner as possible. The model can account for both thermal and electrical contact resistance. This new model will be compared with the existing models using a numerical model as baseline, in order to establish the most accurate analytical model for calculating the performance of a thermoelectric generator.

\section{A model for contact resistance}
\begin{figure}[!t]
  \centering
  \includegraphics[width=0.7\columnwidth]{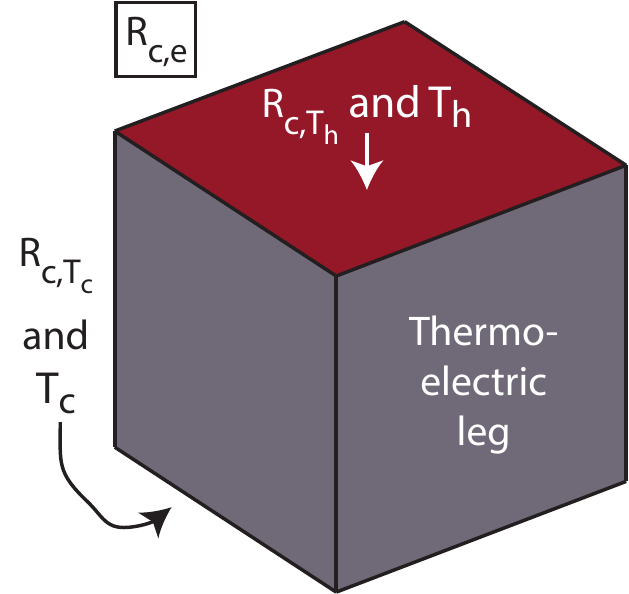}
  \caption{The TE leg investigated. A thermal contact resistance is present at both the hot, $R_\n{c,T_\n{h}}$, and cold side, $R_\n{c,T_\n{c}}$, of the leg. An electrical contact resistance, $R_\n{c,e}$, is also present.}
  \label{Fig_Illustration_single_leg_contact_res}
\end{figure}

We consider a setup with constant hot, $T_\n{h}$, and cold, $T_\n{c}$, side temperature available to the TE leg. This means that the thermal contact resistance will influence only the temperatures that the thermoelectric leg experiences, while the electrical contact resistance will only influence the electrical performance of the leg and not the temperature across the leg. The system considered in shown in Fig. \ref{Fig_Illustration_single_leg_contact_res}.
For a thermal contact resistance, the placement of the contact resistance at either the hot or the cold side of the leg matters, as the thermal contact resistance directly changes the temperature across the leg. Therefore, a thermal contact resistance at the hot and cold sides must be treated separately.

We consider the efficiency, $\eta$, defined as
\begin{eqnarray}\label{Eq.Efficiency_def}
\eta{}=\frac{P}{Q_\n{in}}
\end{eqnarray}
where $P$ is the electrical power produced by the leg at the optimal load resistance, and $Q_\n{in}$ is the heat flowing into the leg.

\subsection{Thermal contact resistance}
We first consider the influence of thermal contact resistances. Consider a thermal resistance on the hot side of the leg of value $R_\n{c,T_\n{h}}$, and a resistance at the cold side of value $R_\n{c,T_\n{c}}$. In general throughout this work the subscript $c$ or $leg$ denotes the contact or leg resistance, respectively, while $e$ or $T$ denotes the electrical or thermal resistance, respectively. As the heat flux is constant, the system can be considered as a thermal resistance circuit. In this circuit, the following equalities hold:
\begin{eqnarray}
\frac{T_\n{h}-T_\n{h,leg}}{R_\n{c,T_\n{h}}} = \frac{T_\n{h,leg}-T_\n{c,leg}}{R_\n{leg,T}} = \frac{T_\n{c,leg} - T_\n{c}}{R_\n{c,T_\n{c}}}
\end{eqnarray}
where $R_\n{leg,T}$ is thermal resistance of the leg and $T_\n{h,leg}$ and $T_\n{c,leg}$ are the actual temperatures experienced by the thermoelectric leg. Solving the set of equations gives the hot and cold side temperatures of the leg as
\begin{eqnarray}\label{Eq.T_c_leg}
T_\n{c,leg} &=& \frac{\frac{R_\n{c,T_\n{h}}}{R_\n{leg,T}}T_\n{c}+\frac{R_\n{c,T_\n{c}}}{R_\n{leg,T}}T_\n{h}+T_\n{c}}{1+\frac{R_\n{c,T_\n{h}}}{R_\n{leg,T}}+\frac{R_\n{c,T_\n{c}}}{R_\n{leg,T}}}\nonumber\\
T_\n{h,leg} &=& \frac{\frac{R_\n{c,T_\n{h}}}{R_\n{leg,T}}T_\n{c}+\frac{R_\n{c,T_\n{c}}}{R_\n{leg,T}}T_\n{h}+T_\n{h}}{1+\frac{R_\n{c,T_\n{h}}}{R_\n{leg,T}}+\frac{R_\n{c,T_\n{c}}}{R_\n{leg,T}}}
\end{eqnarray}
In these expressions, the fraction of the contact resistance to the leg resistance is seen to be the factor controlling the temperature span across the TE leg.


The efficiency of the thermoelectric leg at this new temperature span is then easily calculated using the classical expression for the efficiency of a leg with constant material properties \cite{TE_Handbook_Ch9}
\begin{eqnarray}
\eta = \frac{T_\n{h,leg}-T_\n{c,leg}}{T_\n{h,leg}}\frac{\sqrt{1+Z\bar{T}}-1}{\sqrt{1+Z\bar{T}}+T_\n{c,leg}/T_\n{h,leg}}
\end{eqnarray}
where $\bar{T}=T_\n{c,leg}+(T_\n{h,leg}-T_\n{c,leg})/2$.

\subsection{Electrical contact resistance}
Having established the true temperature span across the thermoelectric leg, the influence of an electrical contact resistance can now be considered. The electrical performance of a thermoelectric leg depends crucially on the electrical resistance external to the leg, i.e. the load resistance. A contact resistance can be considered part of the external electrical resistance, albeit a resistance where the work performed across the resistance cannot be utilized. It matters not if an electrical contact resistance is present on either or both ends of a thermoelectric leg. The electrical contact resistance can be viewed as a single external resistance that influence the thermoelectric leg, regardless of whether it is located at the hot or cold side of the leg, or both. Thus, we consider an electrical circuit with the thermoelectric leg, a contact resistance and a load resistance. The contact resistance is here the sum of the contact resistances on the hot and cold sides of the leg.

The power produced by the thermoelectric leg can be written as a sum of the power dissipated over the contact resistance, $P_\n{c,e}$ and that dissipated over the load resistance, $P_\n{load}$, i.e.
\begin{eqnarray}
P = P_\n{c,e} + P_\n{load} = I^2(R_\n{c,e}+R_\n{load})
\end{eqnarray}
where $I$ is the current and $R_\n{c,e}$ is the electrical contact resistance and $R_\n{load}$ is the load electrical resistance.

We assume that the $I$/$V$-curve of the thermoelectric leg is linear, and that the intersections with the axes are at $V_0$ and $I_0$, respectively. Here $V_0$ is the open circuit voltage and $I_0$ is the closed circuit current. In this case, the current through the circuit can be written in terms of the resistance and $V_0$ and $I_0$ as
\begin{eqnarray}\label{Eq.I}
I = \frac{V_0}{(R_\n{c,e}+R_\n{load})}\frac{1}{\left(1+\frac{V_0}{I_0(R_\n{c,e}+R_\n{load})}\right)}
\end{eqnarray}

At both $V_0$ and $I_0$ the efficiency of the thermoelectric leg is zero. In order to operate the leg at maximum efficiency, the optimal load resistance, $R_\n{load}$, has to be determined. For the case of no contact resistance, the optimal load resistance is equal to
\begin{eqnarray}
R_\n{load,opt} = \frac{V_0}{I_0}
\end{eqnarray}
at which point the power produced is $P_\n{max}=\frac{V_0I_0}{4}=\frac{I_0^2}{4}R_\n{leg,e}$, where $R_\n{leg,e}$ is the electrical resistance of the leg. This is also known as the condition of matched load \cite{TE_Handbook_Ch9}.

When a contact resistance is present, the point of maximum efficiency can be found by considering only the power dissipated over the load resistance, as this is the only factor contributing to the efficiency, i.e.
\begin{eqnarray}\label{Eq.P_load_Rc}
P_\n{load} = I^2R_\n{load} = \frac{V_0^2}{(R_\n{c,e}+R_\n{load})^2}\frac{1}{(1+\frac{V_0}{I_0(R_\n{c,e}+R_\n{load})})^2}R_\n{load}
\end{eqnarray}
where $I$ is taken from Eq. (\ref{Eq.I}). Differentiating this equation with respect to $R_\n{load}$ and equating it to zero, the load resistance at which the power is maximized is found to be at
\begin{eqnarray}
R_\n{load,opt,R_\n{c}} = \frac{V_0}{I_0}+R_\n{c} = R_\n{load,opt}+R_\n{c,e}
\end{eqnarray}
Note that the above equation can be used to determine the contact resistance of a given leg, for which the material properties are known. By using a numerical model to find the optimal load resistance (or by simply computing the resistance of the leg, as the condition of matched load is fulfilled), and by comparing with an experimentally determined optimal load resistance, the difference between these two is simply the contact resistance.

We can express the open circuit voltage, $V_0$, and the closed circuit current, $I_0$, in terms of the power produced in the case of no contact resistance, $P_\n{max}$, as
\begin{eqnarray}
I_0 &=& 4P_\n{max}/V_0 \nonumber\\
V_0 &=& \sqrt{4R_\n{leg,e}P_\n{max}}
\end{eqnarray}

Inserting these into the equation for the power dissipated across the load resistance, Eq. (\ref{Eq.P_load_Rc}), we finally get
\begin{eqnarray}
P_\n{load} = P_\n{max}\frac{R_\n{leg,e}}{R_\n{c,e}+R_\n{leg,e}}=P_\n{max}\frac{1}{1+\delta_\n{e}}
\end{eqnarray}
where $\delta_\n{e}$ is the fraction of the electrical contact resistance to the leg resistance, $\delta_\n{e}=\frac{R_\n{c,e}}{R_\n{leg,e}}$.

In order to find the efficiency, we assume that the heat flux entering the thermoelectric leg is constant, regardless of the value of the external electrical resistance. This assumption will be justified subsequently. As the heat flux remains constant, the decrease in efficiency is easily calculated using Eq. (\ref{Eq.Efficiency_def}). The efficiency is simply decreased by the factor found above, as compared to the optimal efficiency, $\eta_\n{opt}$,
\begin{eqnarray}\label{Eq.factor_Rce}
\eta = \eta_\n{opt}\frac{1}{1+\delta_\n{e}}
\end{eqnarray}

Thus, in total the efficiency of a thermoelectric leg with thermal contact resistances and electrical contact resistances is
\begin{eqnarray}
\eta = \frac{T_\n{h,leg}-T_\n{c,leg}}{T_\n{h,leg}}\frac{\sqrt{1+Z\bar{T}}-1}{\sqrt{1+Z\bar{T}}+T_\n{c,leg}/T_\n{h,leg}}\frac{1}{1+\delta_\n{e}}
\end{eqnarray}
where the hot and cold leg temperatures are given by Eq. (\ref{Eq.T_c_leg}).

\subsection{Alternative analytical models}
The analytical model derived above can be compared with other existing analytical models of TE legs. As mentioned above, two other analytical models that describe the influence of contact resistance on the performance of the thermoelectric device exist.

One model can be constructed by combining the work of Ebling et. al. \cite{Ebling_2010}, with the thermal model presented above, i.e. Eq. (\ref{Eq.T_c_leg}). In the work of Ebling et. al. \cite{Ebling_2010}, the factor for the electrical contact resistance found in Eq. (\ref{Eq.factor_Rce}), is multiplied to the $ZT$ value of the material is order to obtain an effective $ZT$ at the given electrical contact resistance, instead of multiplying it directly to the efficiency.

Another model is the expression for the efficiency of a thermoelectric leg, in the presence of both an electrical and thermal contact resistance, derived by Min and Rowe \cite{Min_1992,Rowe_1996,TE_Handbook_Ch9}. The efficiency is given as
\begin{eqnarray}\label{Eq.Min_Rowe}
\eta = \frac{\Delta{}T}{T_\n{h}}\frac{1}{\left(1+2\frac{R_\n{c,T}}{R_\n{leg,T}}\right)^2\left(2-\frac{1}{2}\frac{\Delta{}T}{T_\n{h}}+\frac{4}{Z T_\n{h}}\frac{1+2R_\n{c,e}/R_\n{leg,e}}{1+2R_\n{c,T}/R_\n{leg,T}}\right)}
\end{eqnarray}
where the subscript nomenclature is as in the rest of this manuscript.\footnote{In the derivation of Eq. (\ref{Eq.Min_Rowe}), we have assumed that there is a typo in the expression given in Ref. \cite{TE_Handbook_Ch9}, Eq. (11.4). A factor of $l_c$ is missing in the last parenthesis in the denominator, i.e. the equation should be $(l+nl_c)/(l+2rl_c)$.}.

It is an interesting fact that for the case of no contact resistance the expression by Min and Rowe reduces to
\begin{eqnarray}
\eta = \frac{\Delta{}T}{T_\n{h}}\frac{1}{2-\frac{1}{2}\frac{\Delta{}T}{T_\n{h}}+\frac{4}{Z T_\n{h}}}
\end{eqnarray}
which is not identical to the classical expression for the efficiency for a leg with constant material properties \cite{TE_Handbook_Ch9},
\begin{eqnarray} \label{Eq.Eta_TE_hand}
\eta = \frac{\Delta{}T}{T_\n{h}}\frac{\sqrt{1+Z\bar{T}}-1}{\sqrt{1+Z\bar{T}}+T_\n{c}/T_\n{h}}
\end{eqnarray}

As previously mentioned in the introduction, these models differ from the current work in several ways. The model by Min \& Rowe \cite{Min_1992,Rowe_1996,TE_Handbook_Ch9}, has previously been shown to be inaccurate for segmented legs \cite{Bjoerk_2015}, which causes speculation as to whether the predictions of this model is also accurate for non-segmented legs. The model of Ebling et. al. \cite{Ebling_2010} assumes a change of the $ZT$ value of a material with contact resistance, which is unphysical.

\section{Application of the model}
In order to determine the validity of the developed model, we have computed the efficiency of 16 individual thermoelectric legs of different materials. A numerical Comsol model, which includes all relevant thermoelectric phenomena, is used to calculate the efficiency of the TE leg with varying contact resistances \cite{Bjoerk_2014}. The model fully accounts for all material parameters, and all as a function of temperature.

For each leg of TE material, the electrical and thermal contact resistance (both hot and cold) were varied independently. The values for the contact resistance were varied as 0\%, 5\%, 10\%, 20\%, 30\%, 40\%, 50\%, 60\%, 70\%, 80\%, 90\% and 95\%, where the values are the fraction of contact resistance of the total combined resistance ($R_\n{c}+R_\n{leg}$). Thus, three parameters were varied independently; the electrical contact resistance, the thermal contact resistance on the hot side of the leg, and the thermal contact resistance on the cold side of the leg. A total of 1728 simulations were conducted for each thermoelectric leg, for 16 different thermoelectric legs.

\subsection{TE materials}
\begin{figure}[!t]
  \centering
  \includegraphics[width=1\columnwidth]{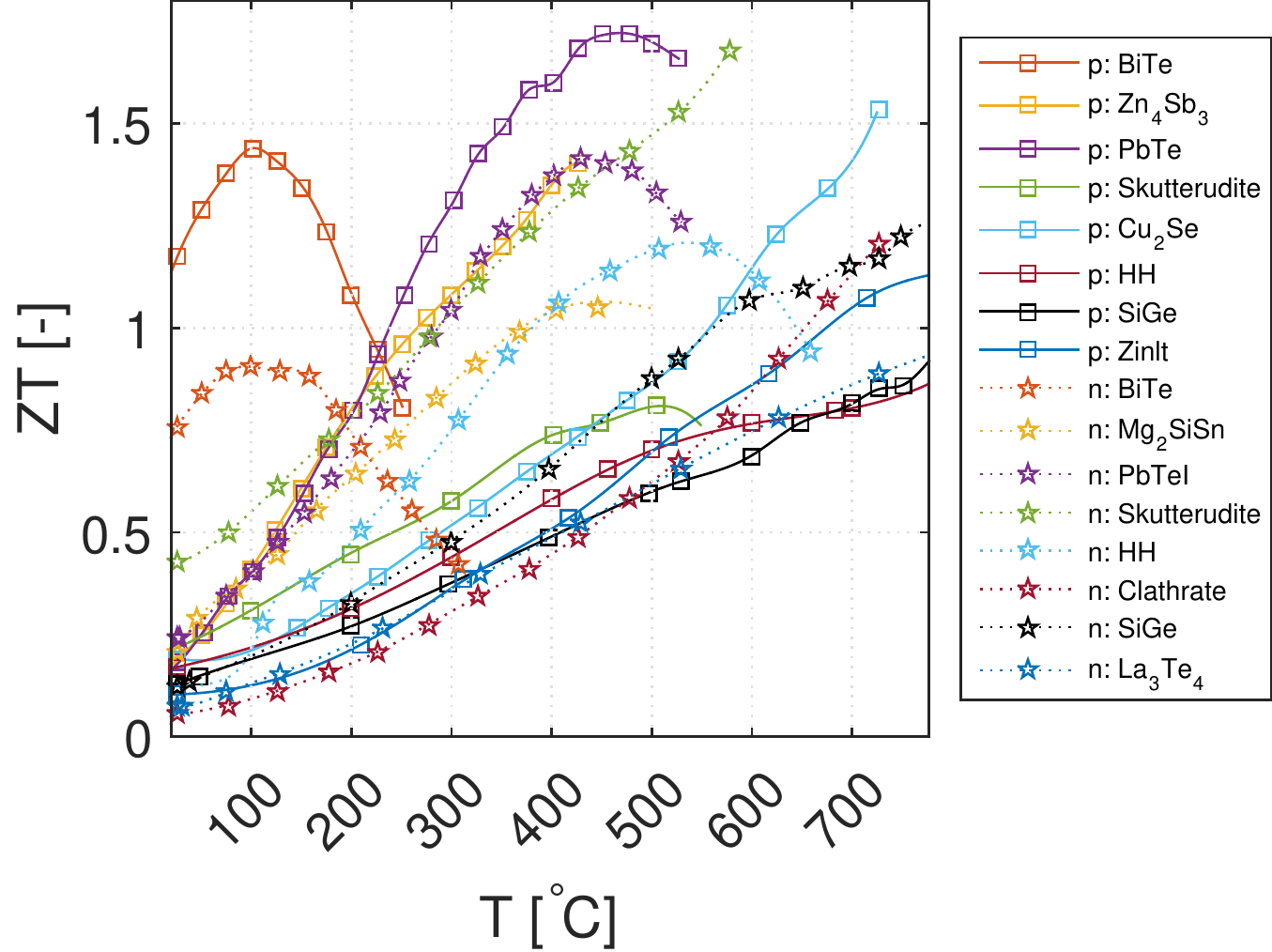}
  \caption{The $ZT$ value of the 16 different thermoelectric materials considered as function of temperature. Only $ZT$ is shown for clarity, but all relevant material parameters have been obtained.}
  \label{Fig_Mat_ZT}
\end{figure}

The materials considered in this study are 8 $p$-type and 8 $n$-type TE materials. The specific $p$-type materials considered are BiSbTe \cite{Ma_2008}, NdFe$_{3.5}$Co$_{0.5}$Sb$_{12}$ (Skutterudite) \cite{Muto_2013}, Yb$_{14}$Mn$_{0.2}$Al$_{0.8}$Sb$_{11}$ (Zinlt) \cite{Toberer_2008}, Zr$_{0.5}$Hf$_{0.5}$CoSb$_{0.8}$Sn$_{0.2}$ (Half-Heusler, HH) \cite{Yan_2010}, PbTe \cite{Pei_2011}, Zn$_4$Sb$_3$ \cite{Chitroub_2008}, Cu$_2$Se \cite{Liu_2012} and SiGe \cite{Joshi_2008}. The specific $n$-type materials considered are BiTe \cite{Kim_2012}, Ti$_{0.5}$Zr$_{0.25}$Hf$_{0.25}$NiSn$_{0.998}$M$_{0.002}$ (Half-Heusler, HH) \cite{Schwall_2013}, Ba$_{8}$Ni$_{0.31}$Zn$_{0.52}$Ga$_{13.06}$Ge$_{32.2}$ (Clathrate) \cite{Shi_2010},\\ Mg$_{2}$Si$_{0.3925}$Sn$_{0.6}$Sb$_{0.0075}$ \cite{Zhang_2008}, PbTe$_{0.9988}$I$_{0.0012}$ \cite{Lalonde_2011},\\ Ba$_{0.08}$La$_{0.05}$Yb$_{0.04}$Co$_{4}$Sb$_{12}$ (Skutterudite) \cite{Shi_2011}, La$_{3}$Te$_{4}$ \cite{May_2010} and SiGe \cite{Wang_2008}. All materials have temperature dependent experimentally measured properties, and the calculated $ZT$ values for the different materials are shown in Fig. \ref{Fig_Mat_ZT}. The materials are identical to those considered by Ref. \cite{Bjoerk_2015} and partly to those considered by Ref. \cite{Ngan_2014}. All material parameters, as a function of temperature, have been obtained from the cited references and are used in the following calculations, but for clarity only $ZT$ is shown in Fig. \ref{Fig_Mat_ZT}. The remaining material properties are available from the author upon request, or are of course available from the cited references.

In all cases, the load resistance was varied to determine the maximum efficiency. The hot side temperature was taken to be close to the peak $ZT$ temperature. The cold side temperature was kept constant at 20 $^\circ{}$C. The geometry of the TE leg matters not, as only the efficiency is considered. The hot side temperature, without contact resistance, and the corresponding efficiency is given in Table \ref{Table.Mats}.

\begin{table}[!b]
\begin{tabular}{c c c S[table-format=3.2] c}
Type & Material  & $T_\n{hot}$   & $\eta$ & Ref.\\ 
     &           & [$^\circ{}$C] & {[\%]} &     \\ \hline
n    & BiTe         & 200   &  6.88    & \cite{Kim_2012} \\
n    & Clathrate    & 700   &  8.01    & \cite{Shi_2010}\\
n    & HH           & 600   &  10.77   & \cite{Schwall_2013} \\
n    & La$_3$Te$_4$ & 1000  &  12.25   & \cite{May_2010} \\
n    & Mg$_2$SiSn   & 500   &  10.97   & \cite{Zhang_2008}\\
n    & PbTeI        & 500   &  11.96   & \cite{Lalonde_2011}\\
n    & SiGe         & 900   &  14.05   & \cite{Wang_2008}\\
n    & Skutterudite & 550   &  14.32   & \cite{Shi_2011} \\
p    & BiTe         & 200   &  6.88    & \cite{Ma_2008} \\
p    & Cu$_2$Se     & 700   &  11.20   & \cite{Liu_2012}\\
p    & HH           & 700   &  9.91    & \cite{Yan_2010} \\
p    & PbTe         & 500   &  13.28   & \cite{Pei_2011} \\
p    & SiGe         & 900   &  11.23   & \cite{Joshi_2008}\\
p    & Skutterudite & 500   &  8.70    & \cite{Muto_2013}\\
p    & Zinlt        & 900   &  11.85   & \cite{Toberer_2008} \\
p    & Zn$_4$Sb$_3$ & 400   &  9.90    & \cite{Chitroub_2008} \\
\end{tabular}
\caption{The $p-$ and $n$-type TE legs considered. The maximum efficiency in percent is given, as well as the hot side temperature. The cold side temperature was kept constant at 20 $^\circ{}$C.} \label{Table.Mats}
\end{table}

\begin{figure*}[!p]
\centering
\subfigure[a]{\includegraphics[width=0.9\columnwidth]{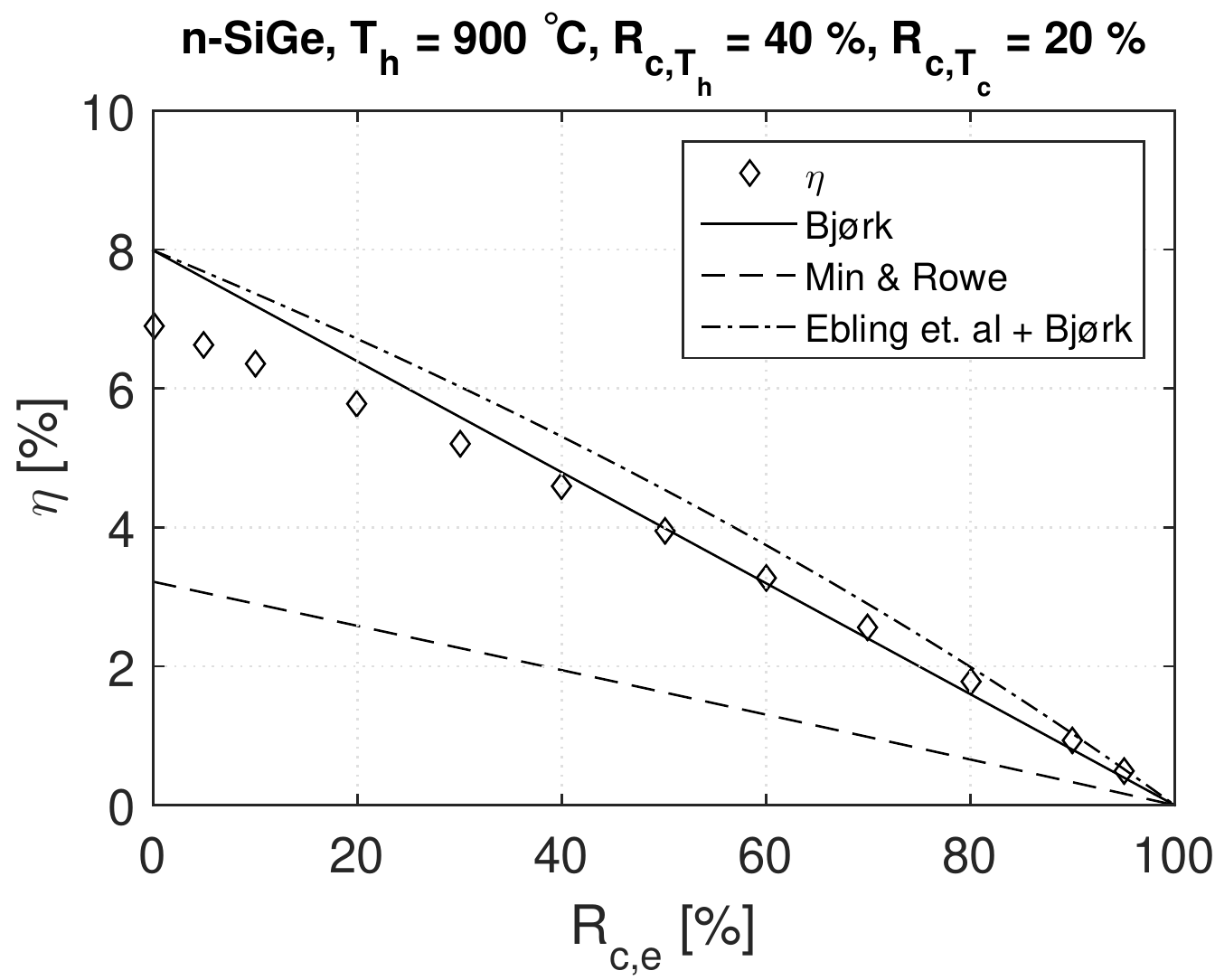}}\hspace{0.2cm}
\subfigure[b]{\includegraphics[width=0.9\columnwidth]{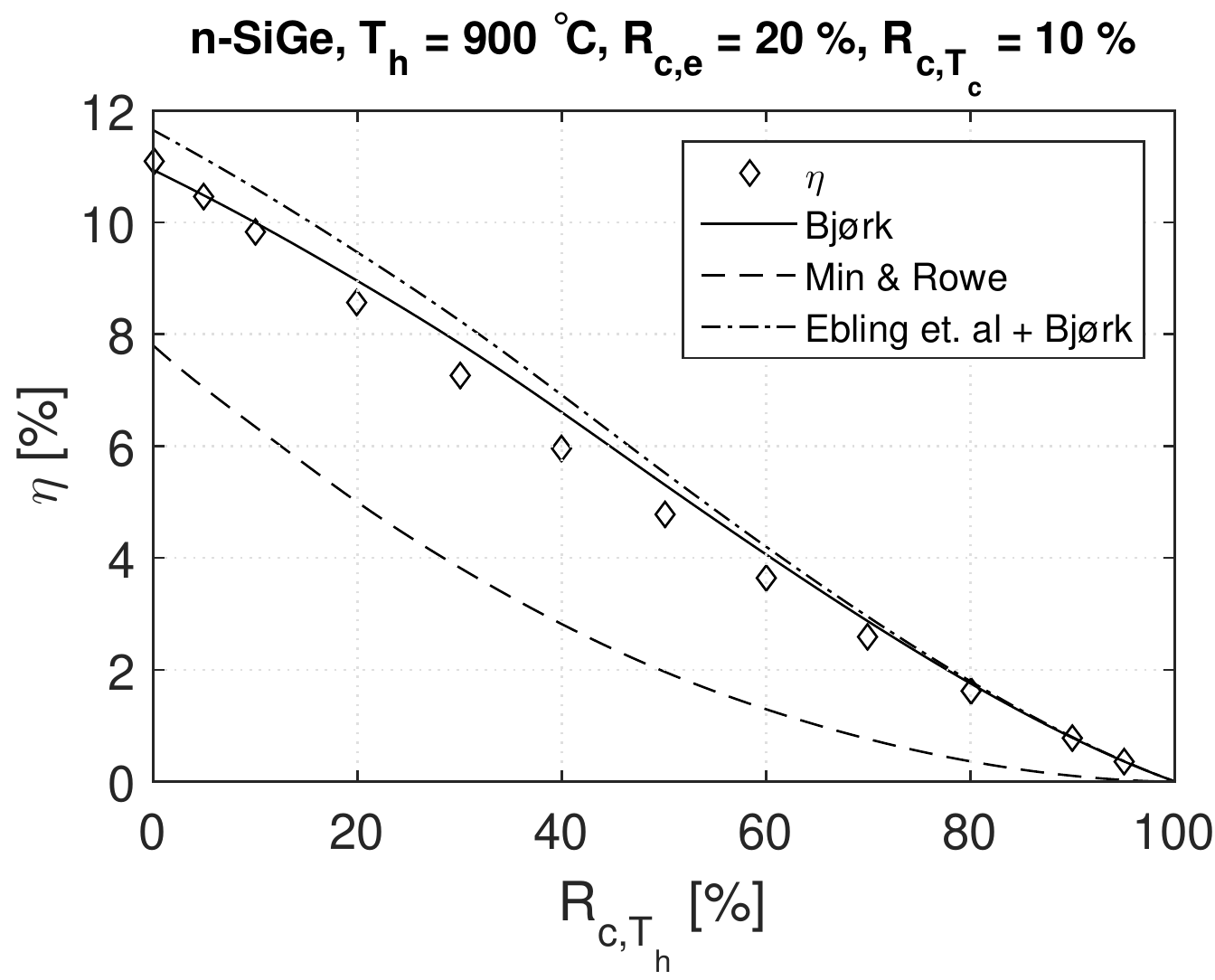}}
\caption{The efficiency as function of contact resistance in the case of a varying (a) electrical or (b) (hot) thermal contact resistance, for the contact resistances given in the figure title. The result of the various analytical models discussed are also shown. The value of the contact resistances are given as percentage of the total combined resistance, i.e. $R_\n{c}/(R_\n{c}+R_\n{leg})$.}\label{Fig.Curves}
\end{figure*}

\begin{figure*}[!p]
\centering
\subfigure[a]{\includegraphics[width=0.9\columnwidth]{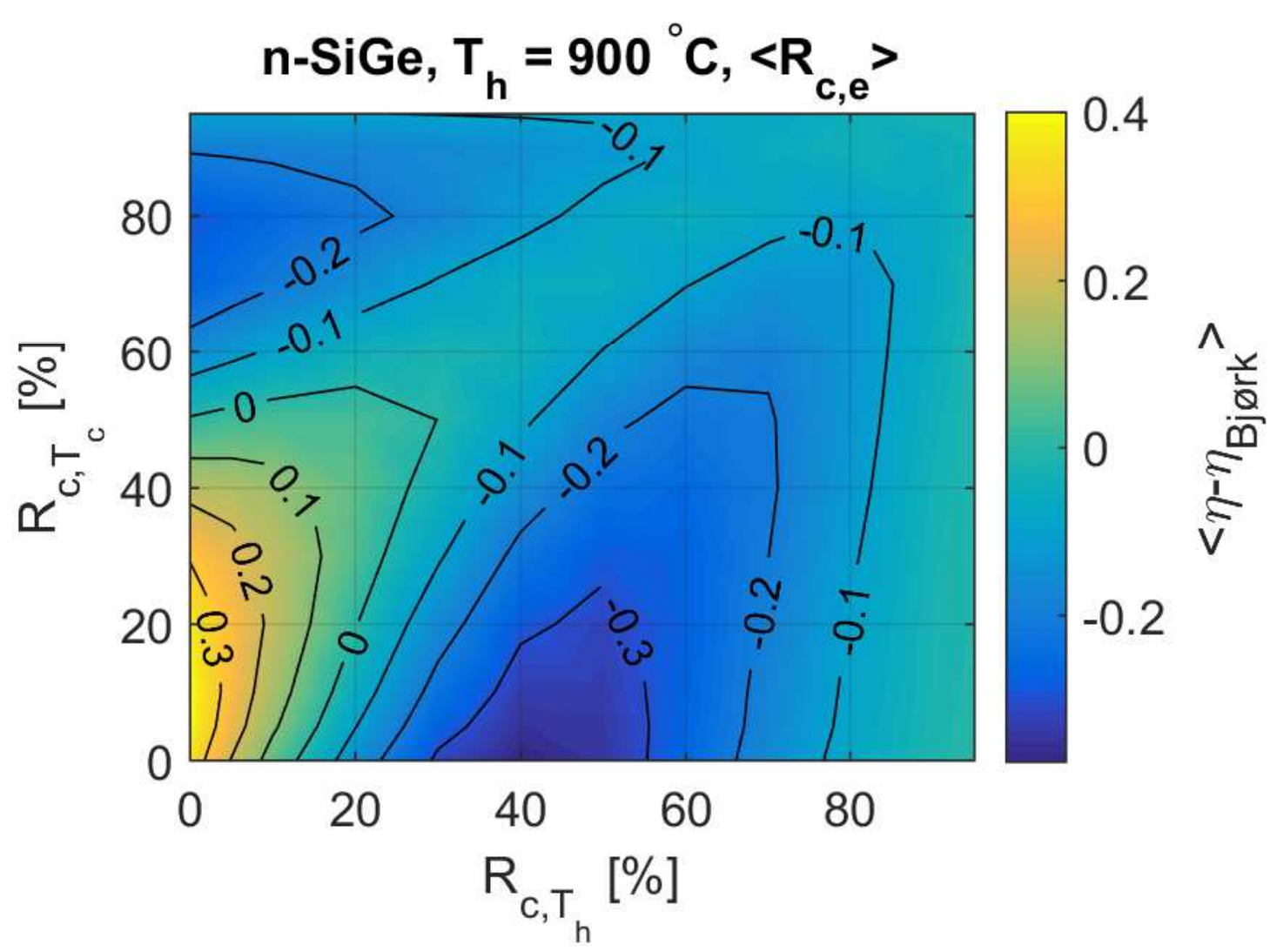}}
\subfigure[b]{\includegraphics[width=0.9\columnwidth]{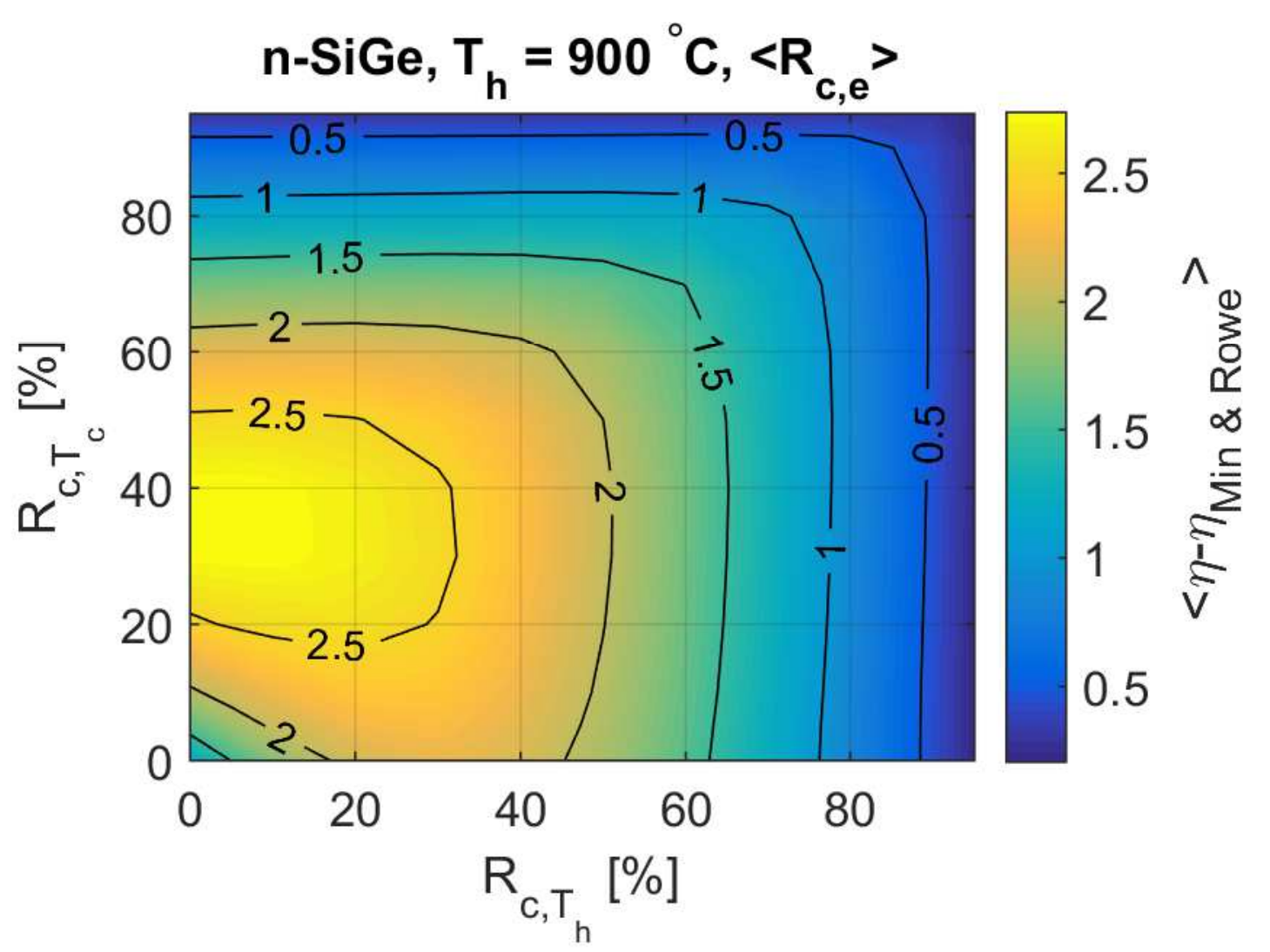}}
\subfigure[c]{\includegraphics[width=0.9\columnwidth]{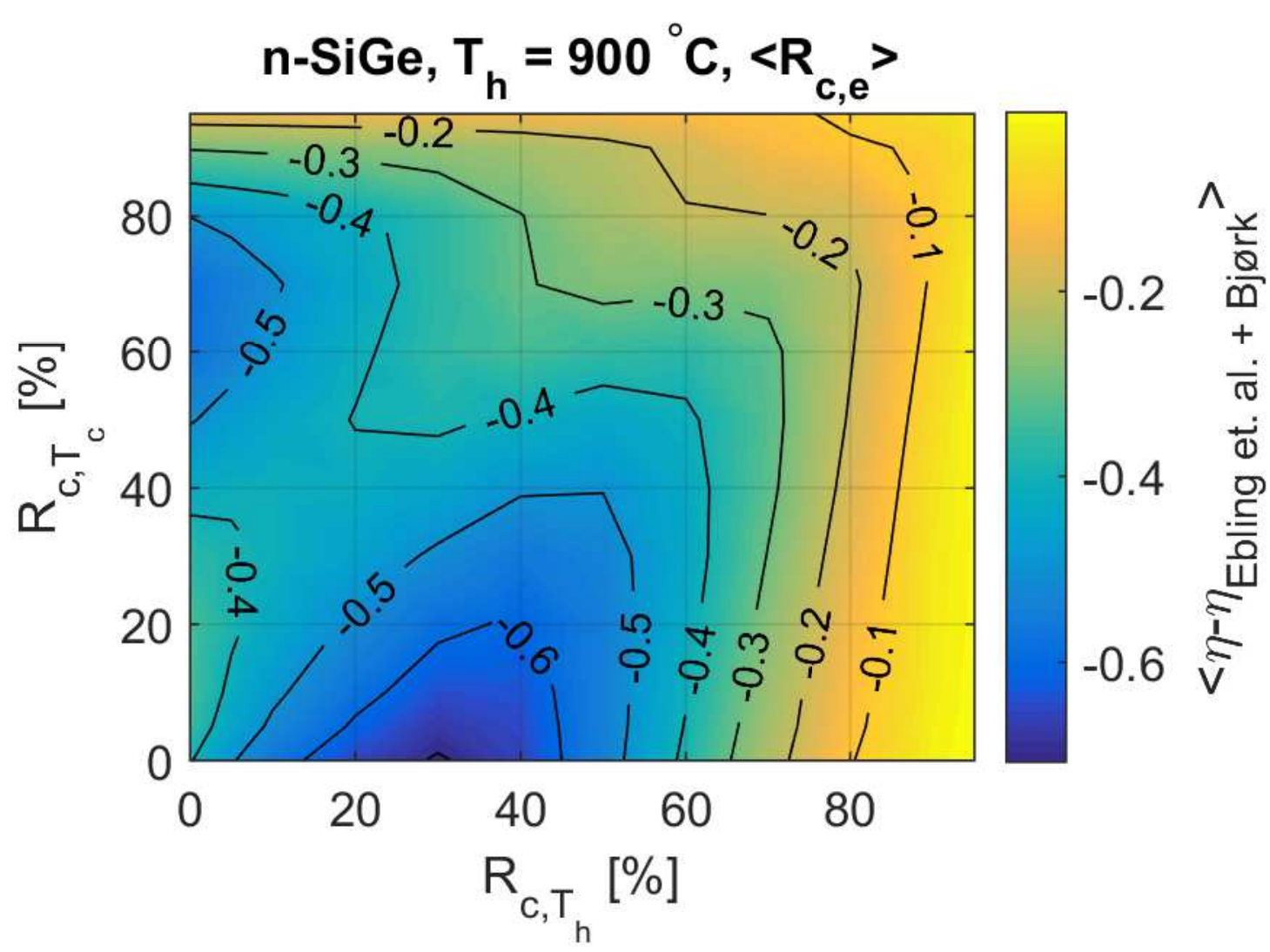}}
\caption{The efficiency as function of the hot and cold thermal contact resistance. The efficiency has been averaged for all values of the electrical contact resistance. The figures show the difference between the computed efficiency and the model prediction of (a) Bj\o{}rk, (b) Min \& Rowe and (c) Ebling et. al. + thermal model from Bj\o{}rk is shown. The value of the contact resistances are given as percentage of the total combined resistance, i.e. $R_\n{c}/(R_\n{c}+R_\n{leg})$.}\label{Fig.surfs}
\end{figure*}

An assumption of the analytical model presented above is that the heat flux does not vary as a function of the external electrical resistance. We have computed this for the 16 materials considered here. For these, the ratio of the heat flux at maximum efficiency and the heat flux at closed circuit current is 0.83 $\pm$ 0.04. As this is the worst case possible, i.e. an infinite electrical contact resistance, the assumption of constant heat flux is justified.

For the analytical model, the $Z$-value is taken to be the average of the $Z$-value across the temperature span actually experienced by the leg, i.e. the temperature span given in Eq. (\ref{Eq.T_c_leg}).

\subsection{Case study: n-SiGe}
As an example, we initially consider the case of $n$-type SiGe. For this material, the hot side temperature is taken to be 900 $^\circ{}$C, at which the efficiency is $\eta = 14.05\%$, in the case of no contact resistances. Shown in Fig. \ref{Fig.Curves} is the computed efficiency and the calculated efficiency from the analytical models, for a case of varying electrical or hot thermal contact resistance. As can be seen from the figure, the prediction of the model presented above (Bj\o{}rk) and the model of Ebling et. al. combined with the thermal model from Bj\o{}rk agree well with the computed efficiency. The model of Min \& Rowe is seen to predict a substantially lower efficiency than is actually the case.

As mentioned above, the efficiency is a function of three independent parameters, namely the electrical contact resistance and the two thermal contact resistances. Therefore, the efficiency as function of these cannot easily be visualized. In Fig. \ref{Fig.surfs} the difference in efficiency between the computed efficiency and that predicted by the different models are shown. The results have been averaged for one type of contact resistance, in order to visualize the results as a surface plot. As can be seen from the figures, the model of Bj\o{}rk and the model of Ebling et. al. combined with the thermal model from Bj\o{}rk predict the efficiency with $\pm 0.6$ pp, while the model of Min \& Rowe predict a too low efficiency.

For the case of $n$-type SiGe, the average difference in efficiency for all contact resistances considered is $-0.08\pm0.43$ pp for the model of Bj\o{}rk, $1.4\pm1.1 $ pp for the case of the model of Min \& Rowe and $-0.32\pm0.35$ pp for the model of Ebling et. al. combined with the thermal model from Bj\o{}rk.

\subsection{All TE materials}
\begin{figure}[!t]
  \centering
  \includegraphics[width=1\columnwidth]{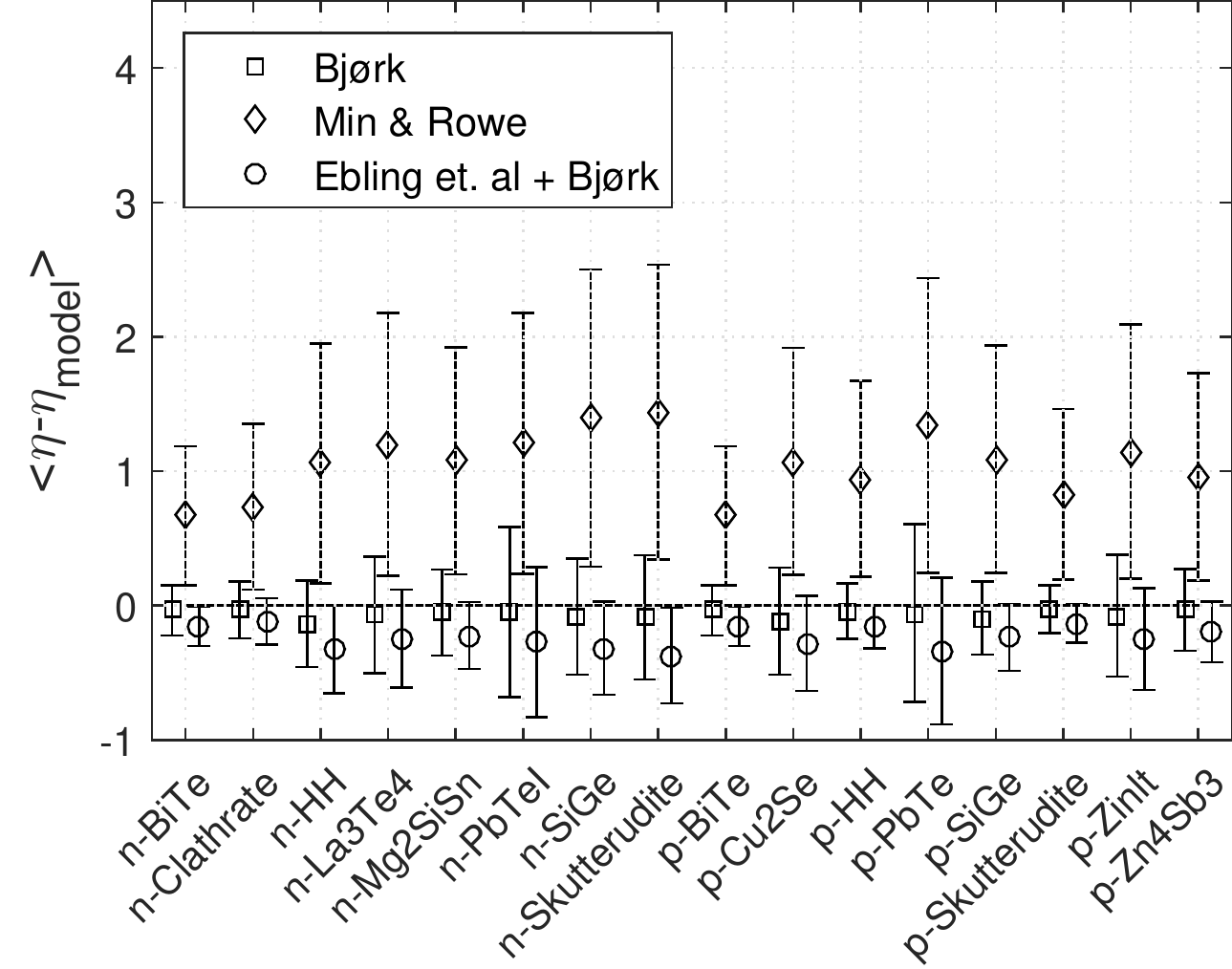}
  \caption{The difference in efficiency for all contact resistances between the computed efficiency and the model by Bj\o{}rk, Min \& Rowe and Ebling et. al. combined with the thermal model from Bj\o{}rk. The errorbars are the standard deviation of the data.}
  \label{Fig_Global_mat_comp}
\end{figure}

The above analysis has been conducted for all materials given in Table \ref{Table.Mats}. Shown in Fig. \ref{Fig_Global_mat_comp} is the average difference between the computed efficiency and the efficiency predicted by the various models, for each material, for all contact resistances. Similar to the conclusion for the specific case of $n$-type SiGe discussed above, the models of Bj\o{}rk and of Ebling et. al. combined with the thermal model from Bj\o{}rk provide the most accurate prediction of the efficiency.

The global average, i.e. for all materials considered here, is a difference in efficiency for the case of the model by Bj\o{}rk of $-0.07\pm0.35$ pp, for the model of Min and Rowe of $1.05\pm0.84$ pp and the model of Ebling et. al. combined with the thermal model from Bj\o{}rk of $-0.24\pm0.30$ pp. Interestingly, the maximum absolute difference in efficiency between the computed efficiency and the models, for all materials, are 1.14 pp for the model of Bj\o{}rk, 3.00 pp for the model of Min and Rowe and 1.17 pp for the model of Ebling et. al. combined with the thermal model from Bj\o{}rk. Thus the first and last models can be trusted to predict the efficiency within at worst 1.2 pp for thermoelectric properties with varying material properties and both thermal and electrical contact resistances, as long as these properties are known.

\section{Conclusion}
An analytical model, capable of calculating the efficiency of a thermoelectric generator with both electrical and hot and cold thermal contact resistances has been presented. The model was compared to a numerical model of a thermoelectric leg, for 16 different thermoelectric materials, as well as to the analytical models of Ebling et. al. and Min \& Rowe. The model presented here was shown to correctly calculate the efficiency for all systems and all contact resistances considered, with a global average difference between the analytical model and the numerical model of $-0.07\pm0.35$ pp. Furthermore, the maximum absolute difference in efficiency between the computed efficiency and the analytical model was 1.14 pp for all materials and all contact resistances considered.

\section*{Acknowledgements}
The author would like to thank the Programme Commission on Sustainable Energy and Environment, The Danish Council for Strategic Research for sponsoring the ``Oxide thermoelectrics for effective power generation from waste heat'' (OTE-POWER) (Project No. 10-093971) project as well as the ``CTEC - Center for Thermoelectric Energy Conversion'' (Project No. 1305-00002B) project. The author also wish to thank the European Commission for sponsoring the ``Nano-carbons for versatile power supply modules'' (NanoCaTe) (FP7-NMP Project No. 604647) project.

\end{document}